\documentclass{ifacconf}

\usepackage{graphicx}      

\usepackage[algo2e,ruled,vlined,linesnumbered]{algorithm2e}

\usepackage[latin1]{inputenc}
\usepackage{amsfonts}
\usepackage{mathtools}
\usepackage{natbib}  
\usepackage{xcolor}  
\usepackage{flushend}

\graphicspath{{./figs/}}

\SetKwRepeat{Do}{do}{while}

\IncMargin{1.5em}

\newtheorem{example}{Example}
\newtheorem{theorem}{Theorem}
\newtheorem{proposition}{Proposition}

\newcommand{\rank}{{\rm rank}}
\newcommand{\tr}{{\rm tr}}
\newcommand{\K}{\mathbb{K}}

\newcommand{\R}{\mathbb{R}}
\newcommand{\F}{\mathbb{F}}
\newcommand{\Pp}{\mathbb{P}}
\newcommand{\X}{\mathcal{X}}
\newcommand{\Null}{{\rm Null}}

\newenvironment{proof}{\begin{trivlist} \item[{ \bf Proof:}] }
	{~\hfill$\Box$ \end{trivlist} }

\begin{document}
	\begin{frontmatter}
		\title{On the Search for Equilibrium Points of Switched Affine Systems}
	
		
		\thanks[footnoteinfo]{This work was supported by the Wallenberg Artificial Intelligence, Autonomous Systems and Software Program (WASP) funded by Knut and Alice Wallenberg Foundation.  The  authors  are grateful  for the  discussions  with  Johan Löfberg.}
		
		\author[First]{Lucas N. Egidio} 
		\author[First]{Anders Hansson} 
		
		\address[First]{Department of Electrical Engineering, Link\"{o}ping University\\
			581 83 Link\"{o}ping, Sweden.\\ {\tt \{lucas.egidio, anders.g.hansson\}@liu.se} }
		
		\begin{abstract}               
			  One of the main aspects of switched affine systems that makes their stabilizability study intricate is the existence of (generally) infinitely many equilibrium points in the state space. Thus, prior to designing the switched control, the user must specify one of these equilibrium points to be the goal or reference. This can be a cumbersome task, especially if this goal is partially given or only defined as a set of constraints. To tackle this issue, in this paper we describe algorithms that can determine whether a given goal is an equilibrium point of the system and also jointly search for equilibrium points and design globally stabilizing switching functions. 
		\end{abstract}
		
		\begin{keyword}
			Switched affine systems, stabilizability, equilibrium points, LMIs.
		\end{keyword}
		
	\end{frontmatter}

\section{Introduction}
Switched affine systems are a particular case of hybrid systems where the dynamic equation is an affine function of the state that is also discontinuous, given the occurrence of switching events. These systems are of great practical interest as they can exactly model the behavior of several power converters, especially dc-dc converters, see~\cite{deaecto2010switched,egidio2020contributions}. However, compared to switched linear systems, their stabilizability problem presents an additional difficulty: the designer has to provide \textit{a priori} a full state vector to be the goal point along with an associated convex combination, see for instance~\cite{bolzern2004quadratic,deaecto2010switched,hetel2014local,sanchez2019practical}. Depending on the system dimension and on the number of subsystems, this can be done by brute-force methods, such as gridding, or by analytically solving a system of nonlinear equations. However, these are no longer viable options when a general case is assumed, which can hinder the versatility and applicability of these methods. This problem becomes even more difficult when the designer is only interested in controlling some output value since these methods require knowledge of the full state equilibrium. 

In this context, this paper presents the following contributions:
\begin{itemize}
	\item The problem of verifying if a given goal point is an equilibrium of a switched affine system is stated and solved using linear programming. Additionally, one algorithm is developed to construct a polytope containing all convex combinations of dynamics associated with this equilibrium point, which can be useful in a subsequent design step.
	\item The problem of finding an equilibrium point satisfying linear constraints on the system output is stated but, in this case, no polynomial-time solution is known as this is an NP-hard problem. In this case, we propose the use of a Genetic Algorithm combined with convex optimization problems to jointly design a stabilizing switching function and find an equilibrium point satisfying design constraints, also adequately bounding a guaranteed cost of performance.
\end{itemize}

To the best of our knowledge, this is the first work devoted to this study and we try to provide details that can serve as a basis for future discoveries. In Section~\ref{sec:probl} we formally state the problem to be studied and in Section~\ref{sec:alg} we characterize the state and output equilibria. In Section~\ref{sec:joint} we investigate whether the switching function design can be combined with the search for a desired equilibrium. Some examples illustrating the discussed theory are presented. Conclusions and directions to future work are provided in Section~\ref{sec:con}.

\textbf{Notation}: $X'$ denotes the transpose of a matrix $X$. A vector of ones with adequate size is $\mathbf{1}$.  The set of $N$ first positive integers is $\K=\{1,\dots,N\}$, the unit simplex of dimension $N$ is $\Lambda=\{\lambda\in\R^N ~:~ \mathbf{1}'\lambda = 1, ~\lambda\geq 0  \}$, and, similarly, we also define the set $\Theta_p =\{\theta\in\R^p~:~\mathbf{1}'\theta = 1 \}$.
\section{Problem statement}\label{sec:probl}
	Consider a switched affine system given by the state-space representation
	\begin{align}
	\dot{x}(t)&=A_{\sigma(t)}x(t)+b_{\sigma(t)},\quad x(0)=x_0\label{eq:sys1}\\
	z(t)&=Cx(t)\label{eq:sys2}
	\end{align}
	where $x(t)\in\R^{n_x}$ is the state variable, $z(t)\in\R^{n_z}$ is an output vector, and $\sigma(t)\in\K$ is a switching signal that selects, at each time instant, one of the available $N$ subsystems given by the pairs $(A_i,b_i),~i\in\K,$ to govern the state evolution. A question that arises in this framework is whether a point $x^*\in\R^{n_x}$ is an equilibrium point of this system for some signal $\sigma^*(t)$, i.e., $A_{\sigma^*(t)}x^*+b_{\sigma^*(t)}=0$ for all $t\geq0$. Trivially, a set of points 
	\begin{equation}
	\X_C \coloneqq \{x\in\R^{n_x}~:~A_ix +b_i = 0,~~i\in\K\}
	\end{equation}
	contains the equilibrium points associated with constant switching signals $\sigma(t)=i,~i\in\K,$ for all $t\geq0$. However, these are somewhat less interesting equilibria for the system~\eqref{eq:sys1}-\eqref{eq:sys2} as they only consider Carath\'{e}odory solutions. Instead, if one allows Filippov solutions~\citep{liberzon2003switching} to occur, another set 
	\begin{equation}
	\X_F \coloneqq\left\{ x\in\R^{n_x}~:~  A(\lambda) x+b(\lambda) =0,~~\lambda\in\Lambda \right\}\label{eq:X_F}
	\end{equation}
	defines the equilibrium points for system~\eqref{eq:sys1}-\eqref{eq:sys2}, where $ A(\lambda) = \sum_{i\in\K} \lambda_iA_i$ and $ b(\lambda) = \sum_{i\in\K} \lambda_ib_i$ and . For each $x^*\in\X_F$ there is at least one $\lambda\in\Lambda$ that satisfies the equality in~\eqref{eq:X_F} and we say that such $\lambda$ is \textit{associated with} $x^*\in\X_F$. Notice that even in cases where $\X_C$ is an empty set, $\X_F$  may have the cardinality of the continuum, as the following example illustrates.
	\begin{example} Consider system~\eqref{eq:sys1}-\eqref{eq:sys2} with $A_1=A_2=0$, ~$b_1=1$, and $b_2=-1$. Allowing only for Carath\'{e}odory solutions, no $x^*\in\R$ satisfies $A_ix^* +b_i = 0$ for any $i\in\K=\{1,2\}$, and thus $\X_C=\emptyset$. However, when Filippov solutions are considered, $\lambda=[0.5~~0.5]'\in\Lambda$ assures that 
	$\X_F =\R$ as $ A(\lambda) x+b(\lambda)=0$ holds for all $x \in \R$.
	\end{example}

	 In this example, the fact that an arbitrary $x^*$ is an equilibrium point is assured by the occurrence of infinitely many switching events within any finite interval of time. It is clear that, in real-world applications, this is a considerably hard requirement to be fulfilled. Nonetheless, this study is intertwined with that of practical stabilizability of switched affine systems with minimum dwell-time constraints, given that in several studies~\citep{deaecto2017stability,sanchez2019practical} the manifold to which the state converges collapses to a point $x^*\in \X_F$ as the dwell-time constraint vanishes.
	 
	 It is important to remark that $x^*\in \X_F$ is only a necessary condition for the existence of a globally stabilizing switching function assuring $x(t)\rightarrow x^*$ as $t\rightarrow \infty$. However, an instance of a sufficient condition for designing this switching function is presented below, adapted from~\cite{deaecto2010switched}
	 \begin{theorem}[\cite{deaecto2010switched}]\label{theo:deacto}
		Consider a switched affine system~\eqref{eq:sys1}-\eqref{eq:sys2} and a given goal $x^*\in\R^{n_x}$.
		For a given matrix $Q>0$, if there exist a symmetric positive definite matrix $P$ and a vector $\lambda\in\Lambda$ satisfying
		\begin{equation}
		A(\lambda)'P+PA(\lambda)<-Q\label{eq:conditions_deaecto_1}
		\end{equation}
		\begin{equation}
		A(\lambda) x^*+b(\lambda)=0\label{eq:conditions_deaecto_2}
		\end{equation}
		 then the switching function $\sigma(t)=u(x(t))$ with
		\begin{equation}
		u(x) = \arg\min_{i\in\K} (x-x^*)'P(A_ix+b_i)
		\end{equation}
		assures the global asymptotic stability of $x^*$ and the upper-bound
		\begin{equation}
		\int_{0}^{\infty} (x(t)-x^*)'Q(x(t)-x^*){\rm d}t \leq (x_0-x^*)'P(x_0-x^*)\label{eq:upperbound_cost_1}
		\end{equation}
	 \end{theorem}
	 
	 This demonstration is presented in~\cite{deaecto2010switched} and, therefore, is omitted. It is important to remark that, when $x^*$ is unknown, both~\eqref{eq:conditions_deaecto_1} and~\eqref{eq:conditions_deaecto_2} present products between $\lambda$ and other design variables, resulting in a non-convex solution set. The same holds for the other previously-mentioned results.

	 The problem we address in this work is the development of an automated procedure to search for a pair $(x^*,\lambda^*)$ such that $x^*\in\X_F$, associated with $\lambda^*\in\Lambda$, satisfy some given design constraints. Also, $\lambda^*$ should assure the existence of $P>0$ fulfilling the design conditions presented in Theorem~\ref{theo:deacto}. Notice that the set $\X_F$ is not necessarily convex, which prevents us from searching for $x^*\in\X_F$ directly by using convex optimization. In the special case where a goal point $x^*$ is given \textit{a priori}, checking whether $x^*\in\X_F$ and obtaining the associated $\lambda\in\Lambda$ can be solved in polynomial-time complexity by linear programming. However, no general algorithm describing how to obtain $\lambda^*\in\Lambda$ from a partially given $x^*$ was documented to the best of our knowledge.

	 In the following section we study particularities of two cases, defined as follows:
	 \begin{enumerate}
	 	\item \textit{State equilibrium}: $x^*$ is completely known \textit{a priori}.
	 	\item \textit{Output equilibrium}: $x^*$ is unknown but subject to 
	 	\begin{equation}
	 	z^*=Cx^*,~~Hx^*\leq g
	 	\end{equation} for given $z^*\in\R^{n_z}$, $H\in\R^{n_x\times l}$ and $g\in\R^l$.
	 \end{enumerate}
	 this study is done regardless of the stability guarantees, which will be tackled in a subsequent section.

	 \section{Equilibrium Characterization}\label{sec:alg}
	 \subsection{State equilibrium}
	 The case where the designer fully knows $x^*$ prior to designing the switching function is an ideal one and, as previously mentioned, can be handled efficiently. However, it carries some properties that are to be discussed herein.
	 
	 The next proposition is the first step towards finding $\lambda^*$ associated with a given $x^*$. But, first, let us define $\tilde{A} = [A_1,\cdots A_N]$, $\tilde{b} = [b_1,\cdots b_N]$, and $M(x^*) = \tilde{A}(\mathbf{1}\otimes x^*)+\tilde{b}$.
	 
	 \begin{proposition} \label{prop:not_in_XF}
	 	Let an instance of system~\eqref{eq:sys1}-\eqref{eq:sys2} and a point $x^*\in\R^{n_x}$ be given. If $\dim\big(\Null(M(x^*))\big)=0$, then $x^*\notin \X_F$.
	 \end{proposition}
 	\begin{proof}
 		This proof directly follows from rewriting the equality that defines $\X_F$ in~\eqref{eq:X_F} as
 		\begin{equation}
 		 A(\lambda) x+b(\lambda) = M(x)\lambda = 0 \label{eq:identity_Mx}
 		\end{equation} 
 		Therefore, if $\dim\big(\Null(M(x^*))\big)=0$, then there is no non-zero $\lambda\in\R^{N}\supset \Lambda$ such that $M(x^*)\lambda = 0$.
 	\end{proof}
 	Although simple, this result represents an efficient tool to discard bad equilibrium-point candidates for system~\eqref{eq:sys1}-\eqref{eq:sys2}. Notice that $M(x^*)$ is a linear map on $x^*$ and will be used throughout this paper to ease the notation with special attention to the identity~\eqref{eq:identity_Mx}. We can now deliver a complementary result in the form of the next proposition.

 	\begin{proposition}\label{theo:x_in_XF}
 		Let an instance of system~\eqref{eq:sys1}-\eqref{eq:sys2} and a point $x^*\in\R^{n_x}$, such that $\dim\big(\Null(M(x^*))\big)=m>0$, be given.
 		The following statements are equivalent:
 		\begin{enumerate}
 			\item[\em (1)] $x^*\in \X_F$
 			\item[\em (2)] There exists a non-negative $v\in\R^N$ such that $v\neq0$ and $M(x^*)v=0$.
 		\end{enumerate}
 	\end{proposition}
 	\begin{proof}
		 {\em (1)} $\implies$ {\em (2)}: Assuming $x^*\in \X_F$, Proposition~\ref{prop:not_in_XF} assures that there should exist a $\lambda^*\in\Lambda$ such that $M(x^*)\lambda^* = 0$. Therefore, $v=\lambda^*\neq0$ is non-negative given the definition of $\Lambda$ and satisfies {\em (2)}.
		
		{\em (2)} $\implies$ {\em (1)}: Now, we assume a non-negative $v\in\R^N$ satisfying {\em (2)} to build $\lambda^*=\eta v$ with $\eta = \mathbf{1}' v>0$. One can immediately verify that $\lambda^*\in\Lambda$. Moreover, 
		\begin{align}
		 A({\lambda^*})x^*+b({\lambda^*})&= M(x^*)\lambda^*\nonumber\\
		&=\eta M(x^*)v \nonumber=0
		\end{align}
		concluding the proof.
 	\end{proof}
 
 	This proposition provides conditions that allow us to efficiently decide whether a given $x^*$ belongs to the set of equilibrium points of the system~\eqref{eq:sys1}-\eqref{eq:sys2} or not. Indeed, condition {\em (2)} can be verified in polynomial-time by solving the feasibility problem
 	\begin{align}
 	 M(x^*)\lambda= 0,~~\lambda \geq 0,~~ \mathbf{1}'\lambda = 1\label{eq:opt_prov_find_lmb}
 	\end{align}
     which assures that $x^*\in\X_F$ associated with $\lambda\in\Lambda$.
 	
 	It is important to remark that the solution $\lambda^*$ need not be unique. Additionally, when there exist two distinct $\lambda^{(1)}$ and $\lambda^{(2)}$ solution to~\eqref{eq:opt_prov_find_lmb}, there also exist a continuum of solutions defined by the linear combination $\lambda(\theta) = \theta \lambda^{(1)}+(1-\theta)\lambda^{(2)}$ with $\theta\in\R$ and $\lambda(\theta)\geq 0$. The following theorem formalizes this observation for the general case.

\begin{theorem}
	For a given $x^*\in\X_F$, consider $p>1$ distinct solutions to~\eqref{eq:opt_prov_find_lmb} given as $\lambda^{(1)},\dots,\lambda^{(p)}$ and defining columns of the matrix $L = [\lambda^{(1)},\dots,\lambda^{(p)}]$. The set 
	\begin{equation}
	\Pp(x^*) = \{ \lambda\in\R^{n_x} ~:~\lambda= L\theta,~\lambda\geq0,~\theta\in\Theta_p\}\label{eq:polytope}
	\end{equation}
	is a polytope of dimension $d=\rank(L)$ and every $\lambda^*\in \Pp(x^*) \subseteq \Lambda $ is also associated with $x^* \in\X_F$. 
\end{theorem} 	
 	
\begin{proof}
	Notice that $L'\mathbf{1} = \mathbf{1}$. From the definition of $\Theta_p$, one can conclude that $\Pp(x^*)\subseteq \Lambda$, as $\lambda\in\Pp(x^*)$ implies that $\lambda\geq 0$ and $\mathbf{1}'\lambda = \mathbf{1}'\theta=1$ for some $\theta \in\Theta_p$. Also, notice that any $\lambda\in\Pp(x^*)$ is such that $M(x^*)\lambda =M(x^*)L\theta$  for some $\theta \in\Theta_p$. The product $M(x^*)L= 0$ is a consequence of the fact that each column of $L$ is a solution to~\eqref{eq:opt_prov_find_lmb}, which implies $M(x^*)L\theta=0$ for all $\theta$ and, therefore, $x^*\in\X_F$. The dimension of $\Pp(x^*)$ is $d=\rank(L)$ as $\Pp(x^*)$ is contained in the subspace spanned by the columns of $L$.
\end{proof}

 	An interesting remark is that the $q$ vertices $\hat\lambda^{(1)},\dots, \hat\lambda^{(q)}$ of the largest polytope $\Pp(x^*)$ associated with some $x^*\in\X_F$ lie all in the boundary of $\Lambda$. This is a consequence of the fact that $\Theta_p$ is an unbounded set and the only condition bounding $\Pp(x^*)$ is $\lambda\geq 0$, the same that bounds $\Lambda$. Finally, notice that a vertex $\hat\lambda$ cannot be written as a convex combination of any two other points $\mu, \nu\in\Pp(x^*)\setminus\{\hat\lambda\}$. A formal proof of this last statement can be found within the proof of Proposition 2.2 of~\cite{ziegler1995lectures}.
 	
 	The enumeration of the vertices of $\Pp(x^*)$ is of great interest given that most stabilizability conditions, for instance~\eqref{eq:conditions_deaecto_1}, depend on a given $\lambda$ associated with $x^*$. Having a set of vectors $\lambda$ fulfilling this  provides a new degree of freedom to these conditions. The task of finding this set reduces to the \textit{vertex enumeration problem} that has been already widely studied in computational geometry, see~\cite{avis1992pivoting} and~\cite{ziegler1995lectures}. However, for the sake of completion, let us present an algorithm that allows us to obtain the vertices of $\Pp(x^*)$ in finite time with simple operations. The main idea behind it relies on the following proposition.
 	
 	\begin{proposition}
 		A vertex $\hat\lambda$ of the polytope $\Pp(x^*)$ defined in~\eqref{eq:polytope} is the unique point in the intersection $\Pp(x^*)\cap \F(\Lambda, R)$ where $\F(\Lambda, R)$ is some face of the simplex $\Lambda$ defined as
 		\begin{equation}
 		\F(\Lambda,R) = \{\lambda\in\Lambda~:~R\lambda= 0 \}
 		\end{equation}
 		with $R\in\R^{r\times N}$, $N>r>0$, being a matrix with $r$ rows forming a subset of the standard basis of $\R^{N}$. 
 	\end{proposition}
	 \begin{proof}
	 	As previously discussed, the vertex $\hat\lambda$ must lie in the boundary of $\Lambda$ and, therefore, belongs to a face of the simplex $\Lambda$. Assume that $\F(\Lambda, R)$ is the lowest-dimensional face of $\Lambda$ containing $\hat\lambda$. If $\F(\Lambda, R)$ is a face of dimension 0, i.e., a vertex of $\Lambda$, then the proof is trivial. Otherwise, $\hat\lambda$ lies in the relative interior of $\F(\Lambda, R)$. The proof follows by contradiction: if there exists a $\lambda\in\Pp(x^*)\cap \F(\Lambda, R)$ such that $\lambda\neq\hat{\lambda}$, there always exist $\epsilon>0$ small enough such that both $\mu(\epsilon)$ and $\mu(-\epsilon)$ also belong to $\F(\Lambda, R)$, where $\mu(\epsilon) = \epsilon\lambda +(1-\epsilon)\hat{\lambda}$. Moreover, notice that  $M(x^*)\mu(\epsilon)= 0$ for any $\epsilon\in\R$ which implies that both $\mu(\epsilon)$ and $\mu(-\epsilon)$ belong to $\Pp(x^*)$ as well. However, this allows us to write $\hat{\lambda}$ as a convex combination of $\mu(\epsilon)$ and $\mu(-\epsilon)$, which leads to a contradiction since $\hat\lambda$ would not be a vertex then.
	 \end{proof}
	
	Based on this result, a method for enumerating the vertices of $\Pp(x^*)$ can look for the unique solutions for~\eqref{eq:opt_prov_find_lmb} inside faces of $\Lambda$, starting from faces with dimension 0 up to $N-1$. However, one need not solve all the $2^{N-1}$ feasibility problems arising from enumerating the faces of $\Lambda$ in general.  Notice that if $\hat\lambda \in\Pp(x^*)\cap \F(\Lambda, R)$ then  $\hat\lambda\in \Pp(x^*)\cap \F(\Lambda, \bar R)$ for any other face $\F(\Lambda, \bar R)$ of $\Lambda$ such that $\F(\Lambda,R)\subset \F(\Lambda,\bar R)$. Hence, if a solution to~\eqref{eq:opt_prov_find_lmb} is found in some face $\F(\Lambda, R)$ of $\Lambda$ it is unnecessary to search for unique solutions in higher dimensional faces containing $\F(\Lambda, R)$. Algorithm~\ref{alg:enum_vertices} explores this idea to construct the set of vertices of $\Pp(x^*)$.
 	
	\begin{algorithm2e}[tb]
	\caption{Enumerating vertices of $\Pp$} \label{alg:enum_vertices}
	
	\KwIn{$M(x^*)$, defined by the system model}
	\KwResult{Set of vertices $\mathcal{V}=\{\hat\lambda^{(1)},\dots, \hat\lambda^{(q)}\}$ }
	\hrule\vspace{0.1cm}
		$\mathcal{R} \gets \{I\} $\;
		$\mathcal{F} \gets \emptyset$\;
		$\mathcal{V} \gets \emptyset $\;
		\While{$\mathcal{R}\neq\emptyset$}{
			$\mathcal{R}_0 \gets \emptyset $\;
			\ForEach{$R\in\mathcal{R}$}{
				\For{$i=1,\dots,{\tt nRows}(R)$}{
					$\mathcal{R}_0 \gets \mathcal{R}_0 \cup \{ {\tt delRow}(R,i)\} $\;
				}
			}
			$\mathcal{R} \gets {\tt prune}(\mathcal{R}_0,\mathcal{F}) $\;
			
			\ForEach{$R\in\mathcal{R}$}{
				$\hat\lambda \gets \lambda ~:~$~\eqref{eq:opt_prov_find_lmb},~ $R \lambda=0$\;
				\If{$v^* \neq 0$ }{ 
					$\mathcal{V} \gets \mathcal{V}  \cup \{\hat\lambda \} $\;
					$\mathcal{F} \gets \mathcal{F} \cup {R}$\;
					$\mathcal{R} \gets \mathcal{R}\setminus\{R\}$\;
				}
			}
		}
 	\Return $\mathcal{V}$\;
 \end{algorithm2e}

  This algorithm works in the following manner. At every iteration of the while-loop in Line~4, it tries to detect undiscovered vertices of $\Pp(x^*)$ to add to the set $\mathcal{V}$.  This is done by solving the feasibility problem~\eqref{eq:opt_prov_find_lmb} with the additional constraint $R\lambda=0$, which allows us to consider only solutions in a face $\F(\Lambda,R)$ of $\Lambda$. Notice that, at a given iteration, $\mathcal{R}$ represents a list of faces $\F(\Lambda,R), R\in\mathcal{R},$ of the same dimension on which no solutions were found so far. We can interpret the rows of $R$ as vertices of $\Lambda$ that are being eliminated from the feasibility problem and $R=I$ eliminates all the vertices initially in Line~1. We also initialize the list of found vertices $\mathcal{V}$ and the correspondent list of faces on which these vertices were found $\mathcal{F}$. Subsequently, Lines~5-8 construct $\mathcal{R}_0$ with matrices $R\in\mathcal{R}$ after all possible single eliminations of rows performed by the operation ${\tt delRow}(R,i)$, which returns $R$ with the $i$-th row eliminated. In Line~9, $\mathcal{R}$ receives a subset of $\mathcal{R}_0$ returned by the operation ${\tt prune}(\mathcal{R}_0,\mathcal{F})$ that removes  every $\bar R\in\mathcal{R}$ corresponding to faces that contain $\F(\Lambda, R)$ for some $R\in\mathcal{F}$. This operation avoids looking for solutions in faces where another solution has already been found. Also, this can be efficiently done by calculating $X = \bar R  R'$ for all  $(\bar R,R)\in\mathcal{R}_0\times\mathcal{F}$ and eliminating $\bar R$ from $\mathcal{R}_0$ if $X$ has no zero rows. It implies that vertices of $\Lambda$ discarded by $\bar R$ are also discarded by $R$ and thus $\F(\Lambda, R)\subset \F(\Lambda,\bar R)$. Finally, the for-loop in Line~10 solves~\eqref{eq:opt_prov_find_lmb} with $R\lambda=0$ for each face represented by $R\in \mathcal{R}$ and, if a solution is found then a unique vertex of $\Pp(x^*)$ is found on the face $\F(\Lambda,R)$. The algorithm terminates when there are no more faces to search for vertices in. The following example illustrates this result.

  \begin{example}
  	Consider the switched affine system~\eqref{eq:sys1}-\eqref{eq:sys2} presented in Example 1 by~\cite{hetel2014local}, defined as
  	\begin{equation}
  	A_1 =A_3 = \begin{bmatrix}
  	0 & 2 \\ 
  	2 & -66
  	\end{bmatrix},~  A_2=A_4=\begin{bmatrix}
  	0 & 2 \\ 
  	2 & 54
  	\end{bmatrix} 
  	\end{equation}
  	 $b_1=b_2=[-360 ~~0]'$ and $b_3=b_4=-b_1$. The goal point is $x^* =0$ and the authors indicate that $\lambda^*=(1/4) \mathbf{1}$ is associated with $x^*\in\X_F$. To verify whether there are other $\lambda\in\Lambda$ associated with $x^*=0$, we used Algorithm ~\ref{alg:enum_vertices} to enumerate the vertices of $\Pp(x^*)$ and we found 4 vertices $\hat\lambda^{(1)}= [0.5 ~0~0.5~0]', ~\hat\lambda^{(2)}= [0.5 ~0~0~0.5]', ~\hat\lambda^{(3)}= [0 ~0.5~0.5~0]'$, and $\hat\lambda^{(4)}= [0 ~0.5~0~0.5]'$. To investigate other possible $\Pp(x^*)$ we present a graphical representation of this polytope for 5 different systems formed by minor modifications of these matrices, namely:
  	 \begin{enumerate}
  	 	\item[$\mathcal{S}_0$]: $b_1=b_2=b_3=b_4=0$ 
  	 	\item[$\mathcal{S}_1$]: No modification
  	 	\item[$\mathcal{S}_2$]: $b_4=0$ 
  	 	\item[$\mathcal{S}_3$]: $b_4=-b_1+\mathbf{1}$
  	 	\item[$\mathcal{S}_4$]: $b_4=-b_1+\mathbf{1}$ and $b_2=b_1+\mathbf{1}$
  	 \end{enumerate}
  	 Graphical representations in the three-dimensional space formed by $\lambda_1,\lambda_2$ and $\lambda_3$ are shown in Figure~\ref{fig:simp} where, for each system $\mathcal{S}$ the respective polytope $\Pp(x^*)$ is drawn in red along with the edges of the unit simplex $\Lambda$ in gray. Notice that for the trivial case $\mathcal{S}_0$ we have a switched linear system and the origin is an equilibrium point associated with any $\lambda\in\Lambda$. For $\mathcal{S}_4$ the polytope $\Pp(x^*)$ only contains one vertex and is, therefore, a singleton. 
  	\begin{figure}[tb]
  		\centering
  		\def\svgwidth{0.5\linewidth}{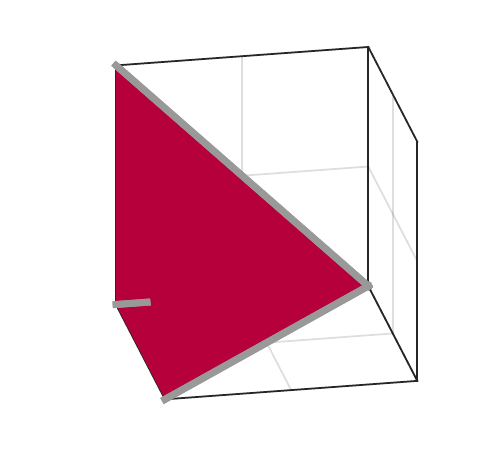}
  		\def\svgwidth{0.5\linewidth}{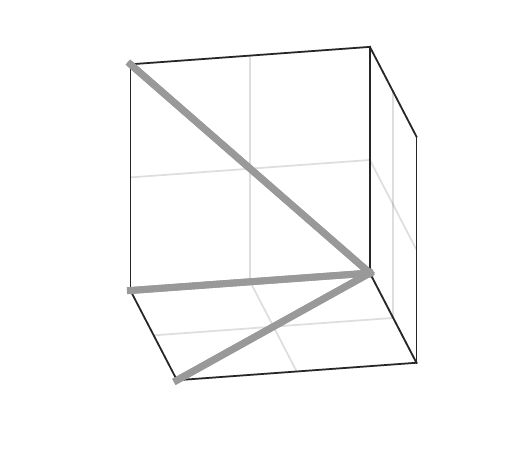}%
  		\def\svgwidth{0.5\linewidth}{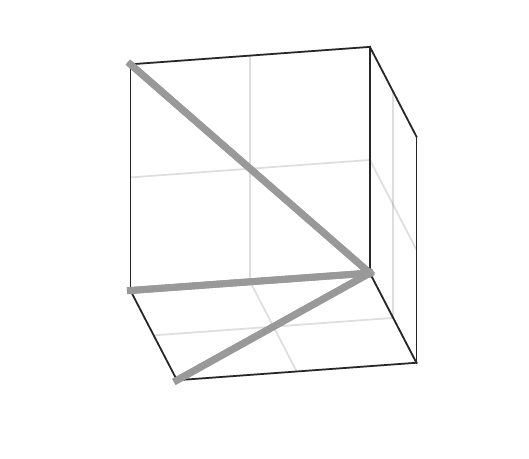}
  		\def\svgwidth{0.5\linewidth}{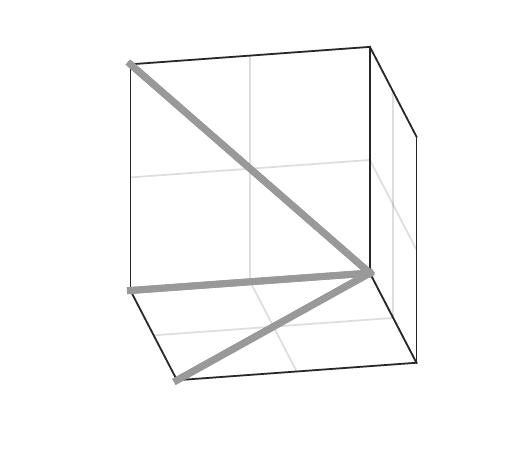}%
  		\def\svgwidth{0.5\linewidth}{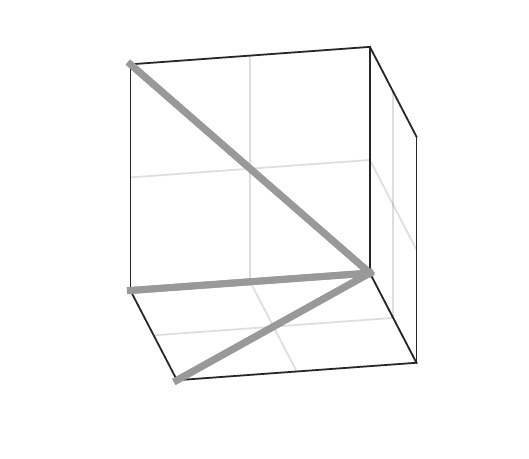}
  		\caption{Graphical representation of the polytope $\Pp(x^*)$ (red) and the unit simplex $\Lambda$ (gray) for different systems.}\label{fig:simp}
  	\end{figure}
  \end{example}

  \subsection{Output equilibrium}
	In contrast with the {state equilibrium} verification problem presented above,  when $x^*$ is a decision variable in the output equilibrium case, the null space $\Null(M(x^*))$ is not defined \textit{a priori}. Therefore, one cannot evaluate the existence of $\lambda\in\Lambda$ associated with some $x^*\in\X_F$ as previously done. Additionally, the set defined by $M(x^*)\lambda=0$ is nonconvex given the product between the unknowns $\lambda$ and $x^*$. Certainly, this problem is more challenging than the one just discussed and it can be cast as a non-linear feasibility problem defined by the constraints 
	\begin{equation}
		M(x^*)\lambda= 0,~~\lambda \geq 0,~~ \mathbf{1}'\lambda = 1,~z^*=Cx^*,~Hx^*\leq g\label{eq:opt_rank_const_2}
	\end{equation}
	where $H$ and $g$ define additional linear design constraints. To the best of our knowledge, no polynomial-time method assured to find a feasible pair $(x^*,\lambda^*)$ is known, in the general case where a solution exists.	In some lower-dimensional instances, an appropriate approach may consist of obtaining the analytical solution that fulfills the conditions in~\eqref{eq:opt_rank_const_2}, see the examples of $\X_F$ provided in~\cite{deaecto2010switched}, for instance. Another alternative is to sample (e.g., in a grid) values in $\Lambda$. However, a methodology for the general case is certainly desirable. One possible approach to this problem is to use local optimization methods to search for feasible solutions. Indeed, one can benefit from the robustness of interior-point methods to potentially find such solutions. Nevertheless, to assure that a solution is found whenever it exists, global optimization methods can be combined with this approach, such as the Branch-and-Bound procedure described in~\cite[p.~187]{tuy2016convex}. 
	
	On the other hand, it is well-known that the higher computational complexity carried by the recursive nature of such global optimization procedure can limit the size of problems to which it is applicable. Thus, one can benefit from stochastic optimization methods to search for feasible solutions, such as evolutionary strategies~\citep{wang2003genetic}. These two approaches will be compared in the next section in the context of finding the pair $(x^*,\lambda)$ while jointly designing the switching function $\sigma(t)=u(x(t))$, as described in Theorem~\ref{theo:deacto}.
	
	\section{Joint Search and Design}\label{sec:joint}

	Given the higher difficulty represented by the non-linear feasibility problem in the search for output equilibria, let us investigate how global optimization methods can be used to determine $x^*\in\X_F$ and $\lambda\in\Lambda$ satisfying~\eqref{eq:opt_rank_const_2}  while satisfying the design conditions for the switching function $\sigma(t) = u(x(t))$ given in Theorem~\ref{theo:deacto}. Even though this section is formulated in the context of output equilibrium, they are readily applied to state-equilibrium searches, which is a particular case of the former.

	This combined approach is useful to jointly design $\sigma(t) = u(x(t))$ and determine the best equilibrium point according to the upper-bound given in~\eqref{eq:upperbound_cost_1}. In other words, by globally solving the optimization problem
	\begin{equation}
	J_2(x^*)\eqqcolon \min_{P,\lambda,x^*}\! (x_0-x^*)'P(x_0-x^*)~\text{s.t.~\eqref{eq:conditions_deaecto_1},~\eqref{eq:opt_rank_const_2}}~~~ \label{eq:output_optimization_prob}
	\end{equation}
	for a given $x_0\in\R^{n_x}$, one can design the switching function in a single-shot without full knowledge of the state equilibrium vector \textit{a priori}. This is possible at the expense of the solution of a more difficult optimization problem. It is important to remark that this is an NP-hard problem as it reduces to verifying whether there exist $\lambda\in\Lambda$ such that $A(\lambda)$ is Hurwitz, see Theorem 1 of~\cite{blondel1997np} as an instance. Despite that, the next example illustrates how this problem can still be handled by readily available branch-and bound solvers in some contexts.
	\begin{example}\label{exa:bnb}
		Consider the switched affine system~\eqref{eq:sys1}-\eqref{eq:sys2} given by matrices
		\begin{equation}
		A_1=\begin{bmatrix}
		-3.1 & 0.3 \\ 
		-0.3 & -2.7
		\end{bmatrix}\!,~b_1=\begin{bmatrix}
		-9 \\ 
		0
		\end{bmatrix}\!,~A_2=\begin{bmatrix}
		-3.2 & 1.1 \\ 
		0.6 & -1.9
		\end{bmatrix}\!,
		\end{equation}
		\begin{equation}
		b_2=\begin{bmatrix}
		-4.5 \\ 
		0.5
		\end{bmatrix}\!,~
		A_3=\begin{bmatrix}
		-8.4 & 0 \\ 
		-2.2 & -3
		\end{bmatrix}\!,~b_3=\begin{bmatrix}
		3.4 \\ 
		-0.2
		\end{bmatrix}\!,
		\end{equation} 
		and $C=[0~~1]$. The goal is to control the output $z(t)=x_2(t)$ to $z^*=0$. Notice that no equilibrium point of any subsystem satisfies this requirement. Solving~\eqref{eq:output_optimization_prob} by using the BMIBNB Branch-and-Bound solver available in YALMIP ~\citep{Lofberg2004} with $Q=I$ and $x_0=\mathbf{1}$ we could obtain  $x^*=[-0.0854 ~~0]'$, $J_2(x^*)=0.2070$,
		\begin{equation}
		P=\begin{bmatrix}
		0.0816 & -0.0209 \\ 
		-0.0209 & 0.1883
		\end{bmatrix},~\lambda= \begin{bmatrix}
		0.3204\\
		0\\
		0.6796
		\end{bmatrix}
		\end{equation}
		This procedure took 6.61 s in Matlab 2019b using an Intel Core i7-8565U with 16 GB of memory\footnote{For more details on the implementation, refer to the source code available in [to be presented].}. In Figure~\ref{fig:bnb_1}, on the top, the state space is represented along with the set of equilibrium points $\X_F$, the found equilibrium point $x^*$ and the line defining the output constraint $Cx=z^*$. On the bottom, the cost $J(x^*)$ evaluated along each point in this line constraint is shown. We can notice that $x^*$ is the global minimizer of $J(x^*)$ on $Cx=z^*$, showing that the Branch-and-Bound method converged to the global optimal solution, as it was expected.
		\begin{figure}
			\def\svgwidth{1\linewidth}{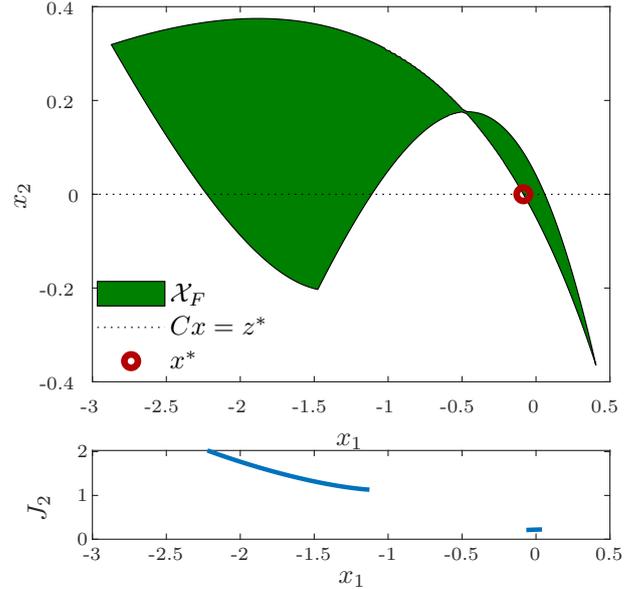}
			\caption{Representation of $\X_F$ and $x^*$ in the state-space (top) and guaranteed cost $J_2$ for equilibrium points on the line $Cx=z^*$ as a function of $x_1$ (bottom).}\label{fig:bnb_1}
		\end{figure}
	\end{example}
	
	The Branch-and-Bound procedure was successful in finding the global optimal solution in this example. However, it may become impracticable for a larger number of subsystems and higher-order systems. One alternative is to employ, for instance, evolutionary strategies to find adequate sub-optimal solutions. We, therefore, propose the use of a genetic algorithm~\citep{wang2003genetic} to determine $\lambda\in\Lambda$ associated with the output requirements. Notice that, given a $\lambda\in\Lambda$ one can very efficiently verify if the correspondent $x^*\in\X_F$ satisfies $Cx^*=z^*$ and $Hx^*\leq g$, and if there exists $P>0$ satisfying the stability condition~\eqref{eq:conditions_deaecto_1}. The main idea of this approach consists in generating populations of vectors $\lambda\in\Lambda$ and apply evolutionary strategies to select those that lead to the best values of $J_2(x^*)$ while satisfying the output constraints. Let us consider that $\lambda$ is provided \textit{a priori}, for a moment. In this case, by adopting an auxiliary variable $\rho>0$, the problem~\eqref{eq:output_optimization_prob} can be rewritten as a convex problem 	
	\begin{equation}
	 \min_{W,~x^*\!,~ \rho} \rho \quad \text{s.t.}~~~\text{\eqref{eq:opt_rank_const_2},}\label{eq:opt_ga}
	\end{equation}
	\begin{equation}
	\begin{bmatrix}
	\rho &(x_0-x^*)'\\
	\bullet & W
	\end{bmatrix}>0,~\begin{bmatrix}
	-WA(\lambda)'-A(\lambda) W & WQ\\
	\bullet & Q
	\end{bmatrix}>0\label{eq:LMIs_ga}
	\end{equation}	
	This is verified by applying the Schur Complement Lemma to the two LMIs in~\eqref{eq:LMIs_ga} and redefining $W=P^{-1}$, which leads to,
	 $\rho>(x_0-x^*)'P(x_0-x^*)$ and $P^{-1}A(\lambda)'+A(\lambda) P^{-1}<-P^{-1}QP^{-1}$, respectively. The first inequality bounds $J(x^*)$ by $\rho$ and the second one, pre- and post-multiplied by $P>0$, is equivalent to~\eqref{eq:conditions_deaecto_1}.

	Therefore, we define the fitness function of an individual $\lambda$ in the population as
	\begin{equation}
	F(\lambda) = \left\{\begin{array}{ll}
	\rho,& \quad\text{if~\eqref{eq:opt_ga}-\eqref{eq:LMIs_ga} is feasible}\\
	\mu + v(\lambda)
	\end{array}  \right.
	\end{equation}
	where $\mu= 10^5$ is a barrier-like term to avoid infeasible solutions and $v(\lambda)$ is the largest residual of the constraints found while trying to solve~\eqref{eq:opt_ga}-\eqref{eq:LMIs_ga} for the given $\lambda\in\Lambda$. This second term is important to allow the strategy to move its candidates towards feasibility. Also, the equality constraint $Cx^*=z^*$ was relaxed by the inequalities $-\epsilon\leq Cx^*-z^*\leq\epsilon$ with $\epsilon=10^{-2}$ to allow for a solution set that can be easier handled numerically.
	
	 The following example illustrates aspects of this strategy in a more challenging problem for which the Branch-and-Bound procedure could not find a feasible solution within a reasonable time. 
	\begin{example}\label{ex:ga}
		Consider the switched affine system defined in~\eqref{eq:sys1}-\eqref{eq:sys2} with
		\begin{align}
		A_1&\!=\!\begin{bmatrix}
		-0.3 & -1.0 & -0.9 \\ 
		0.0 & 1.2 & -1.1 \\ 
		-0.6 & 0.7 & 0.3
		\end{bmatrix},  &
		A_2&\!=\!\begin{bmatrix}
		-0.9 & -1.7 & -0.8 \\ 
		0.8 & 0.7 & 1.0 \\ 
		2.3 & 0.4 & -1.0
		\end{bmatrix}\\
		A_3&\!=\!\begin{bmatrix}
		0.3 & 1.3 & -0.4 \\ 
		1.0 & -2.1 & -0.1 \\ 
		0.5 & -0.3 & 1.0
		\end{bmatrix},  &
		A_4&\!=\!\begin{bmatrix}
		-1.0 & 1.4 & -1.2 \\ 
		0.9 & -0.6 & 0.8 \\ 
		-1.7 & 0.5 & -0.1
		\end{bmatrix}\\
		A_5&\!=\!\begin{bmatrix}
		-0.9 & -1.6 & -0.4 \\ 
		2.0 & 0.8 & 0.5 \\ 
		-0.3 & 0.0 & 0.8
		\end{bmatrix},  &
		A_6&\!=\!\begin{bmatrix}
		-0.7 & -1.4 & -0.2 \\ 
		0.5 & 1.0 & -1.6 \\ 
		0.7 & 0.1 & -0.4
		\end{bmatrix}\\
		A_7&\!=\!\begin{bmatrix}
		0.8 & -0.2 & -0.6 \\ 
		1.6 & 0.2 & 0.0 \\ 
		1.2 & 0.4 & -0.7
		\end{bmatrix},  &
		A_8&\!=\!\begin{bmatrix}
		-0.8 & -0.5 & 0.0 \\ 
		1.4 & -0.8 & 0.2 \\ 
		0.9 & 1.4 & 0.2
		\end{bmatrix}
		\end{align}
		\begin{align}
		b_1\!&=\!\begin{bmatrix}
		-0.7 \\ 
		-1.2 \\ 
		-0.2
		\end{bmatrix}\!\!,&\!\!\!\!
		b_2\!&=\!\begin{bmatrix}
		-1.9 \\ 
		-1.1 \\ 
		0.6
		\end{bmatrix}\! \!,&\!\!\!\!
		b_3\!&=\!\begin{bmatrix}
		-0.7 \\ 
		0.3 \\ 
		-1.9
		\end{bmatrix} \!\!,&\!\!\!\!
		b_4\!&=\!\begin{bmatrix}
		1.3 \\ 
		-0.5 \\ 
		-0.7
		\end{bmatrix}\! \!,\\
		b_5\!&=\!\begin{bmatrix}
		-0.6 \\ 
		0.1 \\ 
		-0.3
		\end{bmatrix} \!\!,&\!\!\!\!
		b_6\!&=\!\begin{bmatrix}
		-0.1 \\ 
		0.1 \\ 
		-0.4
		\end{bmatrix} \!\!,&\!\!\!\!
		b_7\!&=\!\begin{bmatrix}
		0.1 \\ 
		1.6 \\ 
		-0.2
		\end{bmatrix} \!\!,&\!\!\!\!
		b_8\!&=\!\begin{bmatrix}
		6.6 \\ 
		0.9 \\ 
		4.1
		\end{bmatrix}\!\!,
		\end{align}
	and $C=[0 ~0~ 1]$. Notice that all subsystems are unstable. However, we want to find $\lambda\in\Lambda$ associated with $x^*\in\X_F$ such that $z^*=0$, $x^*$ is stabilizable and the upper bound $J(x^*)$ for the quadratic cost~\eqref{eq:upperbound_cost_1} is minimized, with $Q=I$ and $x_0=\mathbf{1}$. The Branch-and-Bound procedure adopted in Example~\ref{exa:bnb} was unable to find the global solution before a timeout of 3600 s occurred. The first feasible solution was found after approximately 1410 s and the best value found for the objective function was $J(x^*) = 8.5349$.	
	 On the other hand, using the Genetic Algorithm \texttt{ga()} available in the Global Optimization Toolbox in Matlab with a population of $1000$ individuals provided a feasible solution at the end of the eighth iteration (i.e., in 1209.54 s) and converged to a local optimum defined by 
	\begin{equation}
	\lambda= \begin{bmatrix}
	0.0602 \\ 
	0.1571 \\ 
	0.1205 \\ 
	0.1096 \\ 
	0.1011 \\ 
	0.2866 \\ 
	0.0866 \\ 
	0.0793
	\end{bmatrix},~P=\begin{bmatrix}
	3.6875 & 1.4173 & 0.1474 \\ 
	1.4173 & 3.3900 & -0.7493 \\ 
	0.1474 & -0.7493 & 5.4987
	\end{bmatrix},
	\end{equation} 
	$x^* = [-0.0351~ 0.2721~  0.01]$
	 corresponding to $J(x^*) = 12.4950$.  In Figure~\ref{fig:ga_fig}, on top, the best value found for each generation of the Genetic Algorithm is depicted. The entire procedure took 4161.61 s in the same setup used for Example~\ref{exa:bnb} but employing parallelism over 8 CPU cores to solve~\eqref{eq:opt_ga}-\eqref{eq:LMIs_ga}. To investigate the reliability of this approach, this test was repeated 100 times with different seeds for the random number generator and the execution time elapsed until a feasible solution was found was registered for each repetition. Their distribution is illustrated by the histogram in the bottom part of Figure~\ref{fig:ga_fig}, which shows that the Genetic Algorithm found feasible solutions in all repetitions, mostly in less than 800 s.
	
	\begin{figure}
		\def\svgwidth{1\linewidth}{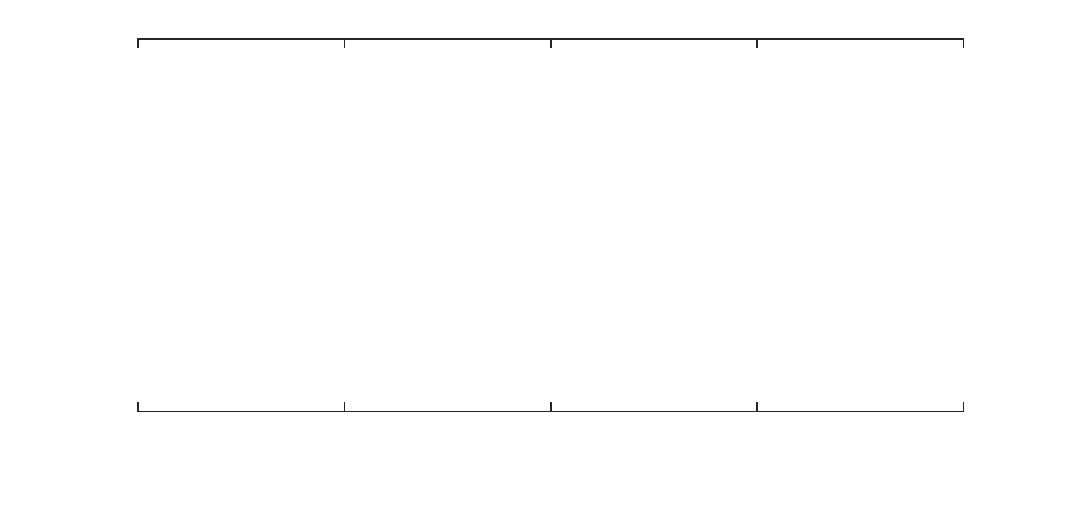}
		\def\svgwidth{1\linewidth}{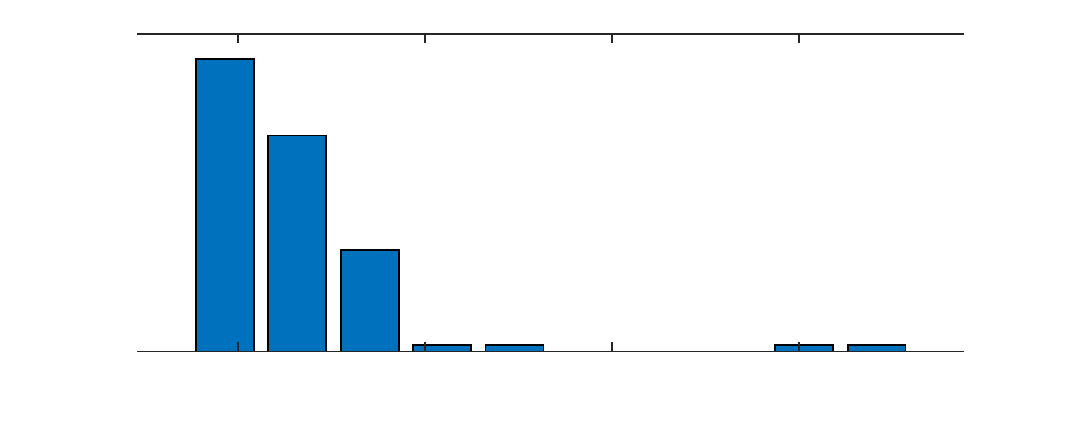}
		\caption{Best value of $\rho$ found at each iteration (generation) of a single genetic algorithm execution (top) and frequency of elapsed time until a feasible solution is found for 100 different executions (bottom).}\label{fig:ga_fig}
	\end{figure}
	\end{example}
	
	One important remark regarding this example is that, by applying suitable cuts to the search space, (e.g., adding a new constraint $\tr(W)< 10^2$), we noticed an increased performance of the Branch-and-Bound method. A methodology for choosing these cuts is, however, outside the scope of this work.
	
\section{Conclusion and Future Work}\label{sec:con}
	We have formulated the problem of searching for vectors $\lambda\in\Lambda$ associated with given and partially given goal points $x^*\in\X_F$ for a switched affine system control design. Two methodologies for the joint search and design of a globally stabilizing switching function are presented and compared. The first one is a Branch-and-Bound procedure which can reliably find solutions whenever they exist but the associated time-complexity may limit its applicability to smaller instances, as in Example~\ref{exa:bnb}. Example~\ref{ex:ga} illustrates how a Genetic Algorithm can be used to find feasible solutions to larger problems faster, in average, than the Branch-and-Bound method. For future work, we propose the development of different approaches that are competitive with these methods, for example, a hybrid methodology combining Genetic Algorithms to initialize Branch-and-Bound solvers or \textit{vice-versa}.

 	\bibliography{equilibria_efficient}

\begin{thebibliography}{13}
\providecommand{\natexlab}[1]{#1}
\providecommand{\url}[1]{\texttt{#1}}
\providecommand{\urlprefix}{URL }
\expandafter\ifx\csname urlstyle\endcsname\relax
  \providecommand{\doi}[1]{doi:\discretionary{}{}{}#1}\else
  \providecommand{\doi}{doi:\discretionary{}{}{}\begingroup
  \urlstyle{rm}\Url}\fi

\bibitem[{Avis and Fukuda(1992)}]{avis1992pivoting}
Avis, D. and Fukuda, K. (1992).
\newblock A pivoting algorithm for convex hulls and vertex enumeration of
  arrangements and polyhedra.
\newblock \emph{Discrete \& Computational Geometry}, 8(3), 295--313.

\bibitem[{Blondel and Tsitsiklis(1997)}]{blondel1997np}
Blondel, V. and Tsitsiklis, J.N. (1997).
\newblock Np-hardness of some linear control design problems.
\newblock \emph{SIAM Journal on Control and Optimization}, 35(6), 2118--2127.

\bibitem[{Bolzern and Spinelli(2004)}]{bolzern2004quadratic}
Bolzern, P. and Spinelli, W. (2004).
\newblock Quadratic stabilization of a switched affine system about a
  nonequilibrium point.
\newblock In \emph{IEEE American Control Conference}, volume~5, 3890--3895.

\bibitem[{Deaecto et~al.(2010)Deaecto, Geromel, Garcia, and
  Pomilio}]{deaecto2010switched}
Deaecto, G.S., Geromel, J.C., Garcia, F.S., and Pomilio, J.A. (2010).
\newblock Switched affine systems control design with application to {DC}--{DC}
  converters.
\newblock \emph{IET Control. Theory Appl.}, 4(7), 1201--1210.

\bibitem[{Deaecto and Geromel(2017)}]{deaecto2017stability}
Deaecto, G.S. and Geromel, J.C. (2017).
\newblock Stability analysis and control design of discrete-time switched
  affine systems.
\newblock \emph{IEEE Trans. Automat. Contr.}, 62(8), 4058--4065.

\bibitem[{Egidio(2020)}]{egidio2020contributions}
Egidio, L.N. (2020).
\newblock \emph{Contributions to Switched Affine Systems Control Theory with
  Applications in Power Electronics}.
\newblock Ph.D. thesis, University of Campinas.

\bibitem[{Hetel and Bernuau(2014)}]{hetel2014local}
Hetel, L. and Bernuau, E. (2014).
\newblock Local stabilization of switched affine systems.
\newblock \emph{IEEE Trans. Automat. Contr.}, 60(4), 1158--1163.

\bibitem[{Liberzon(2003)}]{liberzon2003switching}
Liberzon, D. (2003).
\newblock \emph{Switching in Systems and Control}.
\newblock Birkh\"auser Boston, Boston.

\bibitem[{L{\"{o}}fberg(2004)}]{Lofberg2004}
L{\"{o}}fberg, J. (2004).
\newblock Yalmip : A toolbox for modeling and optimization in matlab.
\newblock In \emph{In Proceedings of the CACSD Conference}. Taipei, Taiwan.

\bibitem[{Sanchez et~al.(2019)Sanchez, Garcia, Hadjeras, Heemels, and
  Zaccarian}]{sanchez2019practical}
Sanchez, C.A., Garcia, G., Hadjeras, S., Heemels, W., and Zaccarian, L. (2019).
\newblock Practical stabilization of switched affine systems with dwell-time
  guarantees.
\newblock \emph{IEEE Trans. Autom. Control}, 64(11), 4811--4817.

\bibitem[{Tuy(2016)}]{tuy2016convex}
Tuy, H. (2016).
\newblock \emph{Convex analysis and global optimization}.
\newblock Springer.

\bibitem[{Wang(2003)}]{wang2003genetic}
Wang, S.C. (2003).
\newblock Genetic algorithm.
\newblock In \emph{Interdisciplinary Computing in Java Programming}, 101--116.
  Springer.

\bibitem[{Ziegler(1995)}]{ziegler1995lectures}
Ziegler, G.M. (1995).
\newblock \emph{Lectures on polytopes}, volume 152.
\newblock Springer Science \& Business Media.

\end{thebibliography}

\end{document}